# HYBRID DATA MINING TECHNIQUE FOR KNOWLEDGE DISCOVERY FROM ENGINEERING MATERIALS DATA SETS

Doreswamy[1], Hemanth K S[2]

[1,2]Department of Computer Science
Mangalore University, Mangalagangotri-574 199, Karnataka, INDIA,
e-mail: doreswamyh@yahoo.com, reachhemanthmca@gmail.com

## ABSTRACT

*Studying materials informatics from a data mining perspective can be beneficial for manufacturing and other industrial engineering applications. Predictive data mining technique and machine learning algorithm are combined to design a knowledge discovery system for the selection of engineering materials that meet the design specifications. Predictive method-Naive Bayesian classifier and Machine learning Algorithm - Pearson correlation coefficient method were implemented respectively for materials classification and selection. The knowledge extracted from the engineering materials data sets is proposed for effective decision making in advanced engineering materials design applications.*

## 1. INTRODUCTION

The rapid developments in materials science and information technologies have influenced the large volume of massive data sets and materials informatics respectively. Materials informatics a field of study that applies the principles of informatics to materials science and engineering to better understand the use, selection, development, and discovery of materials.

As a lot of traditional analytic techniques employed for materials structural-properties analysis and not effective any longer under these situations, researchers in the manufacturing industries and other industrial engineering applications areas are being faced new research issues in systematic analysis of materials data sets. Therefore, materials informatics has been emerging in material science and technology as a new research areas[5],[10],[11], and has already changed the experimental methods and way of thinking in materials research, and will lead even more challenges in interdisciplinary research.

Data Mining is an interdisciplinary field merging ideas from statistics, machine learning, information science, visualization and other disciplines[7]. It is a very useful approach to integrate information and theory for knowledge discovery from any informatics such Bioinformatics, Chemoinformatics, Nano informatics, Materials informatics and so on. The impact of Data Mining and knowledge discovery has been evidenced by many successful research experimental results[19],[20],[21],[22]. Therefore, Data mining can be used to extract non-trivial, hidden, previously unknown, potential useful and ultimately understandable knowledge from massive materials databases[29],[30]. A typical knowledge discovery process is shown in the figure 1.

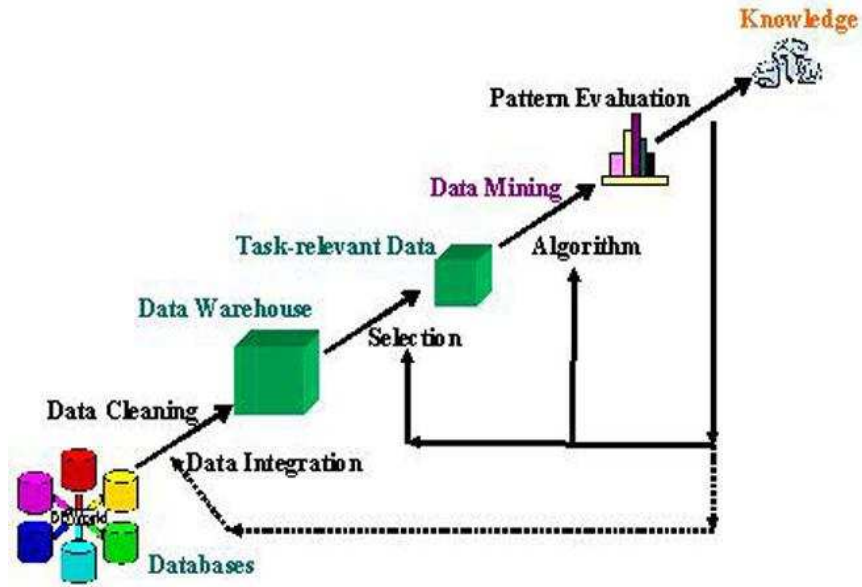

**Fig 1**: Knowledge Discovery Process

Data Mining has two primary Models: Descriptive Data Mining Model and predictive Data Mining Model. Descriptive mining models describe or summarize the general characteristics or behaviour of the data in the materials database. Predictive models perform inference on the current data in order to make the prediction. Both of them are fundamentals to understand materials behaviours. In general , in materials informatics, Data mining can be used in the following task[13]:

(i). **Association analysis:** Association analysis is good at discovering patterns, and can be used to develop heuristic rules for materials behaviour based on large data sets[26],[27],[28].

(ii). **Classifier/Predict modelling:** Some machine learning algorithms can be used for materials class prediction, and materials classification models such as Support Vector Regression (SVR) and Neural Network (NN), can be used to build up the Predict models. These models can be used to predict crystal structure or composite materials properties from fused materials data[14].

(iii). **Cluster analysis:** As an exploratory data analysis tool, it can sort different materials or properties into groups in such a way that the degree of association between two objects is maximal if they belong to the same group and minimal otherwise. And, cluster analysis can be integrated with high-throughput experimentation for rapidly screening combinatorial data[20].

(iv). **Outlier Analysis:** In properties analysis or combinatorial experiments, outlier analysis is used to identify anomalies, especially to assess the uncertainty and accuracy of results, and distinguish between true discoveries and false-positive results.

(v). **Material visualization:** Reconstruction of material structure information based on materials data would help researchers to analyze the relationships between material structure and material properties[16].

The rest of the paper is organized as follows: scope of knowledge discovery on materials informatics and materials data base is discussed in section 2. The section 3 describes naive Bayesian classifier algorithm and Person Correlation measures. The experimental results are presented in the section 4. The conclusions and future scopes are given in the section 5.

## 2. SCOPE OF DATA MINING IN MATERIALS INFORMATICS

Data Mining is becoming an increasingly valuable tool in the broad area of materials development and design[4],[9], and there are good reasons why this area is particularly rich for materials informatics[14],[15],[16],[17]. There is a massive range of possible new materials, and it is often complex to physically model the relationships between constituents, processing and final properties. Therefore, materials are primarily still developed by quantitative and trial-and-error procedures, where researchers are guided by experience and heuristic rules for materials classification, selection and property predictions. These rules are applied to somewhat limited materials data sets of constituents and processing conditions, but then try as many combinations as possible to find materials with desired properties. This is essentially human Data Mining, where one's brain, rather than the computer, is being used to find correlations, make predictions, and design optimal strategies. Transferring Data Mining tasks from human to computer offers the potential to enhance accuracy, handle more data, and to allow wider dissemination of accrued knowledge[5].

Materials informatics[17],[18],[19] has been a subject of materials science, since the international conference of "Materials Informatics-Effective Data Management for New Materials Discovery" was held in Boston in 1999. Wei[24] described that materials informatics is a new subject that leverages information technology and computer network technology to represent, parse, store, manage and analyze the material data, in order to realize the sharing and knowledge mining of materials data for uncovering the essence of materials, and accelerate the new material discovery and design. The research areas of materials informatics are mainly focused on following tasks[25]:

i. **Data standards:** There are thousands of materials databases in different formats[32], and they are difficult to communicate with each other. To standard these databases and to integrate materials data into a single or coherent database, data pre-processing is the first important task of materials informatics[23] to enable knowledge discovery.

ii. **Organization and management of material data:** In order to meet materials researchers' different needs, satisfy the need of research and production, and to construct the materials data sets into a whole and single coherent database, efficient Materials Database Management Systems(MDBMS) is very necessary[25].

iii. **Data mining on materials data:** There is an enormous range of possible new materials, and it is often difficult to physically model the relationships between constituents, and processing, and final properties. Data mining has the abilities for selecting, exploring and modelling large amounts of data to uncover previously unknown patterns from large materials databases[7]. Data mining involves some high-effective computational algorithms[18],[19],[25] such as neural networks, genetic algorithm, fuzzy algorithms and etc.

### 2.1 Material Database

Nowadays, modern science and technology is advancing rapidly and powerfully increasing. Accordingly, materials science and engineering change with each passing day, and new ideas, methods, techniques, processes, and materials appear one after another[5]. So for that region data mining is becoming one of the increasingly valuable tool in the general area of materials development. There is an enormous range of possible new materials, and it is often difficult to physically model the relationships between constituents, and processing, and final properties. As researchers who are researching the new materials, many kinds of material properties database are necessary because the researcher cannot investigate all the properties of all materials in one handbook and it take a lot of time and cost to refer the relevant bibliographic materials. However, there are a few material property databases that are available. Materials database is an organized collection of materials data sets. Each data set characterizes an engineering material with their properties. It is being frequently accessible during concurrent materials design. Therefore, materials database with 5600 materials data sets is designed and organized by referring many textbooks, handbooks and website of engineering materials. The materials database contains all kinds of materials such as polymer, ceramic and metals and other reinforcement materials. Each material data has maximum twenty five sub-properties that belong to different materials properties. The object oriented materials database organized and knowledge discovery process is shown figure 2. Object oriented programming concepts such as inheritance and polymorphism has been utilized in the presented approach. Owing to this design of OODB, an efficient classification task has been achieved by utilizing simple SQL queries.

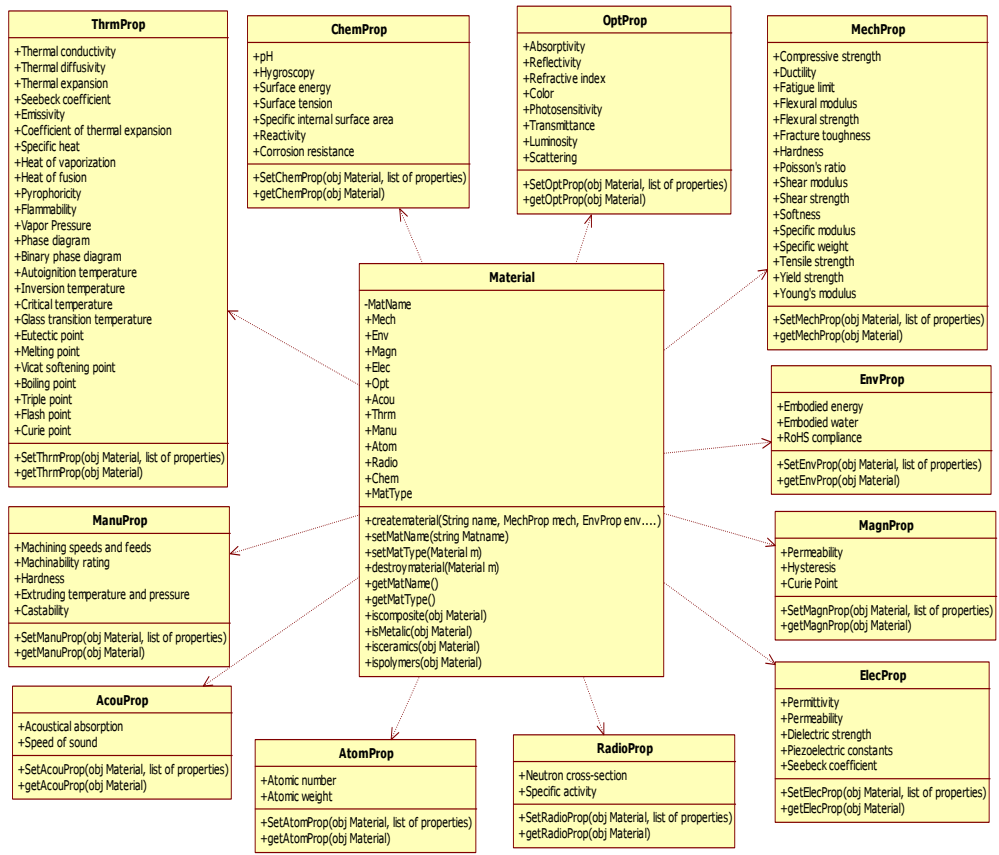

**Fig2: Object oriented engineering material database.**

## 3. DATA MINING TECHNIQUE

Classification and prediction is one of the core tasks of Data Mining. A classification technique is a systematic approach to building classification models from training and testing data sets. Several classification models such as Decision Tree Classifier, Rule-Based Classifier, Neural Network Classifier, naive Bayesian Classifier, Neuro-Fuzzy classifier, Support Vector Machines and etc. are reported in literature. Each technique employs a learning algorithm to identify a model that best fits the relationships between the attribute set and class label of the input data. The model generated by a learning algorithm should both fit the input data well and correctly predict the class labels of data set that has never seen before. Therefore, the key objective of the data mining algorithm is to build models with good generalization capacity.

### 3.1 Naive Bayesian Classifier

Naive Bayesian classifier is a statistical classifier that can predict class membership probabilities such as the probability that a given tuple belongs to a particular class. It is fast and incremental that can deal with discrete and continuous attributes and has excellent performance in real-life problems. The naive Bayesian classifier, or simple Bayesian classifier generally used for classification or prediction task. As it is simple, robust and generality, this procedure has been deployed for various applications such as Materials damage detection[1],[2], Agricultural land soils classification[3], Network intrusion detection[8], Machine learning applications[19]. Therefore, the application of this method is extended to classification of engineering materials data sets[4],[6],[9] and to reduce the computational cost of classification of materials for materials selection. The naive Bayesian classifier procedure is described in the following section.

### 3.2 Algorithm of Naïve Bayesian Classifier

1. Let D be a training set of tuples and their associated class labels. Each tuple is represented by an n-dimensional attribute vector, $X = (x_1, x_2, \ldots x_n)$, depicting n measurements made on the tuple from n attributes, respectively, $A_1, A_2, \ldots A_n$.

2. Suppose that there are m classes, $C_1, C_2, \ldots C_m$. Given a tuple, X, the classifier will predict that X belongs to the class having the highest posterior probability, conditioned on X. That is, the naïve Bayesian classifier predicts that tuple X belongs to the class $C_i$, if and only if

$$P(C_i/X) > P(C_j/X) \text{ for all } 1 \leq j \leq m; j \neq i. \quad (1)$$

3. Thus it maximizes $P(C_i/X)$. The class $C_i$ for which $P(C_i/X)$ is maximized is called the maximum posteriori hypothesis. By Bayes' theorem.

$$P(C_i/X) = \frac{P(X/C_i)P(C_i)}{P(X)} \quad (2)$$

As $P(X)$ is constant for all classes, only $P(X/C_i)P(C_i)$ need be maximized. If the class prior probabilities are not known, then it is commonly assumed that the classes are equally likely, that is, $P(C_1) = P(C_2) = P(C_3) = \ldots = P(C_m)$, and it would therefore maximize $P(X/C_i)$. Otherwise, it maximizes $P(X/C_i)P(C_i)$. Note that the class prior probabilities may be estimated by $P(C_i) = |C_i, D|/|D|$, where $|C_i, D|$ is the number of training tuples of class $C_i$ in D.

4. Given data sets with many attributes, it would be extremely computationally expensive to compute $P(X/C_i)$. In order to reduce computation in evaluating $P(X/C_i)$, the naive assumption of class conditional independence is made. This presumes that the values of the attributes are conditionally independent of one another, given the class label of the tuple (i.e., that there are no dependence relationships among the attributes). Thus,

$$P(X/C) = \prod_{k=1}^{n} P(X_k/C_i)$$
$$= P(X_1/C_i) \times P(X_2/C_i) \times P(X_2/C_i) \times \ldots P(X_n/C_i) \quad (3)$$

The probabilities $P(X_1/C_i), P(X_2/C_i), P(X_3/C_i) \ldots\ldots P(X_n/C_i)$ can easily be estimated from the training tuples. Recall that here $x_k$, refers to the value of attribute $A_k$, for tuple X. For each attribute, the attribute value may be either categorical or continuous-valued. For instance, to compute $P(X/C_i)$, it is considered the following:

If $A_k$ is categorical, then $P(X_k/C_i)$ is the number of tuples of class $C_i$ in D having the value for $A_k$, divided by $|C_i, D|$, the number of tuples of class $C_i$ in D.

5. In order to predict the class label of X, $P(X/C_i)P(C_i)$ is evaluated for each class $C_i$. The classifier predicts that the class label of tuple X is the class $C_i$ if and only if

$$P(X/C_i)P(C_i) > P(X/C_j)P(C_j) \text{ for all } 1 \leq j \leq m; j \neq i \tag{4}$$

In other words, the predicted class label is the class $C_i$ for which $P(X/C_i)P(C_i)$ is the maximum.

### 3.3 Person Correlation Measure

Correlation is one of the most widely used similarity measures in machine learning filed[7]. However, compared with proposed numerous discriminate learning algorithms in distance metric space, only a very little work has been conducted in materials informatics using correlation similarity measure.

Pearson correlation coefficient, $r$ is as follows:

$$r = \frac{\sum XY - \frac{(\sum X)(\sum Y)}{n}}{\sqrt{\left(\sum X^2 - \frac{(\sum X)^2}{n}\right)\left(\sum Y^2 - \frac{(\sum Y)^2}{n}\right)}} \tag{5}$$

Where $n$ is the number of tuples, X and Y are $i^{th}$ attributes, $\sum XY$ is the sum of the cross-product of X and Y. $\sum x$ is the sum of user inputs, $\sum Y$ is the sum of attribute Y, one polymer material, $\sum X^2$ is the sum of squared of attribute values of X and $\sum Y^2$ is the sum of squared of attribute values of Y.

Correlation coefficient, r, value ranges from -1.00 to +1.00. A negative correlation coefficient value $-1.00 < r < 0$ tells that there is a perfect negative similarity between the materials data sets. This means that as values on one material data set increase there is a perfectly predictable decrease in values on the other material data set. In other words, as values of one material data goes up, the other goes in the opposite direction (it goes down). A correlation coefficient value $0 < r < +1.00$ tells you that there is a perfect positive similarity between the material data sets. This means that as values on one material data set increase there is a perfectly predictable increase in values on the other material data set. In other words, as one material data set goes up so does the other, a correlation coefficient of 0.0 tells you that there is a zero correlation, or no similarity, between the material data sets. The experimental results of naive Bayesian classifier and Pearson correlation are discussed in detail in the following section.

## 4. EXPERIMENTAL RESULTS AND DISCUSSION

In this experiment, materials database is organized by sampling material data sets from peer-reviewed research papers published[23],[29] and from poplar materials website http://www.matweb.com as mentioned in above. The tuples of data table consists of both numerical and categorical attribute values. The tuples consisting only categorical attributes and their values are considered for finding probable class of materials. A typical set of training sample data sets is shown in the following table 1.

**Table 1**: Training samples with categorical attributes for Materials classifications.

| CR | CH | CE | SM | CAST | EXTRN | MANFT | CS | MACHN | FS | WA | Class Label |
|---|---|---|---|---|---|---|---|---|---|---|---|
| Excellent | Poor | NIL | Good | Fair | Good | Excellent | Poor | Good | Poor | Poor | P |
| Good | Poor | NIL | Good | Fair | Good | Excellent | Poor | Good | Poor | Poor | P |
| Good | Poor | NIL | Good | Fair | Good | Excellent | Poor | Good | Poor | Poor | P |
| Good | Poor | NIL | Good | Fair | Good | Excellent | Poor | Good | Poor | Poor | P |
| Very Good | Poor | NIL | Good | Fair | Good | Excellent | Poor | Good | Poor | Poor | P |
| Excellent | Poor | Good | Poor | Poor | Poor | Good | Excellent | Poor | Good | Poor | C |
| Excellent | Poor | Good | Poor | Poor | Poor | Good | Excellent | Poor | Good | Poor | C |
| Good | Poor | Good | Poor | Poor | Poor | Good | Excellent | Poor | Good | Poor | C |
| Good | Fair | Good | Poor | Poor | Poor | Good | Excellent | Poor | Good | Poor | C |
| Good | Fair | Good | Poor | Poor | Poor | Good | Excellent | Poor | Good | Poor | C |
| Poor | Very Good | Excellent | Excellent | Excellent | Excellent | Fair | Good | Good | Good | Poor | M |
| Poor | Good | Excellent | Excellent | Excellent | Excellent | Fair | Good | Good | Good | Poor | M |
| Good | Good | Excellent | Excellent | Excellent | Excellent | Fair | Good | Good | Good | Poor | M |
| Fair | Good | Excellent | Excellent | Excellent | Excellent | Fair | Good | Good | Good | Poor | M |
| Poor | Good | Excellent | Excellent | Excellent | Poor | Fair | Good | Good | Excellent | Fair | M |
| Good | Fair | Good | Poor | Poor | Fair | Good | Excellent | Poor | Good | Fair | C |
| Good | Fair | Good | Poor | Poor | Fair | Good | Excellent | Poor | Good | Fair | C |
| Poor | Very Good | Good | Excellent | Excellent | Very Good | Fair | Good | Good | Good | Good | M |
| Poor | Good | Good | Excellent | Excellent | Good | Fair | Good | Good | Poor | Good | M |
| Good | Poor | Good | Good | Fair | Good | Excellent | Poor | Good | Poor | Fair | P |

CR: Chemical Resistance, CH: Conductivity-Heat, CE: Conductivity-Electricity SM: Sheet Metal, CAST: Casting, EXTRN: Extrusion, MANFT: Manufacturing, CS: Creep Strength, MACHN: Mach inability, FS: Fatigue Strength, WA: Water Absorptions

A prototype software module realizing naive Bayesian classifier is designed and developed using .NET technology as it is efficient for handling object oriented data objects. This software module accepts design requirements from the user's Graphical User Interface(GUI) and predicts probable class to which design requirements belong. The design requirements associated the geometrical features of the materials are determined by design engineers or through CAD systems. The GUI of the implemented software module is shown in the following figure 3.

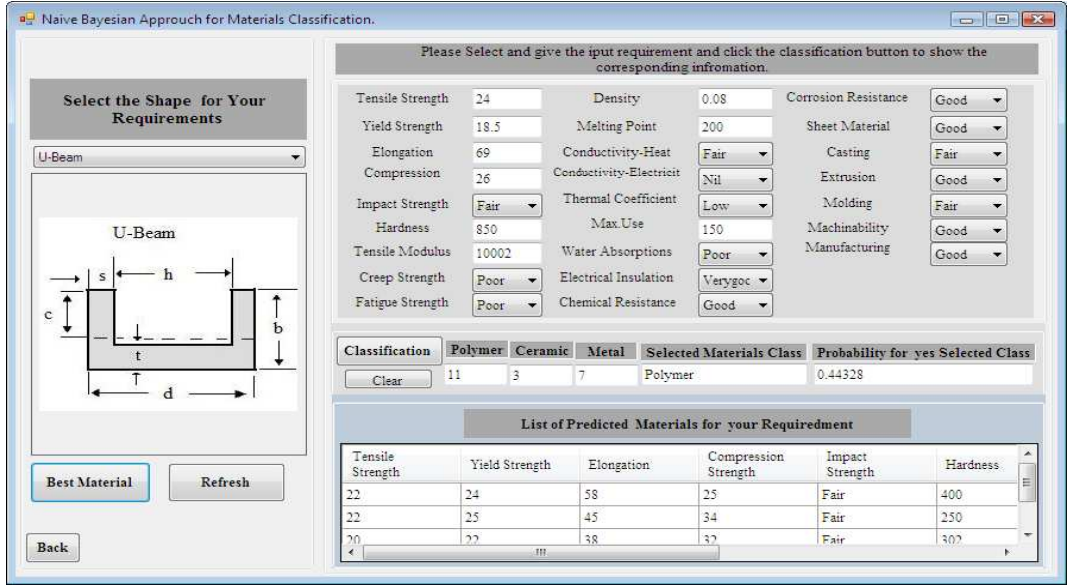

**Fig 3:** A prototype software module for probable material's class prediction on input design requirements.

After finding the input design requirements class, the materials data sets in the materials data base that belonging to the input design requirements class are clustered. Computation of degree of similarity between two materials data set is shown in following the prototype software module.

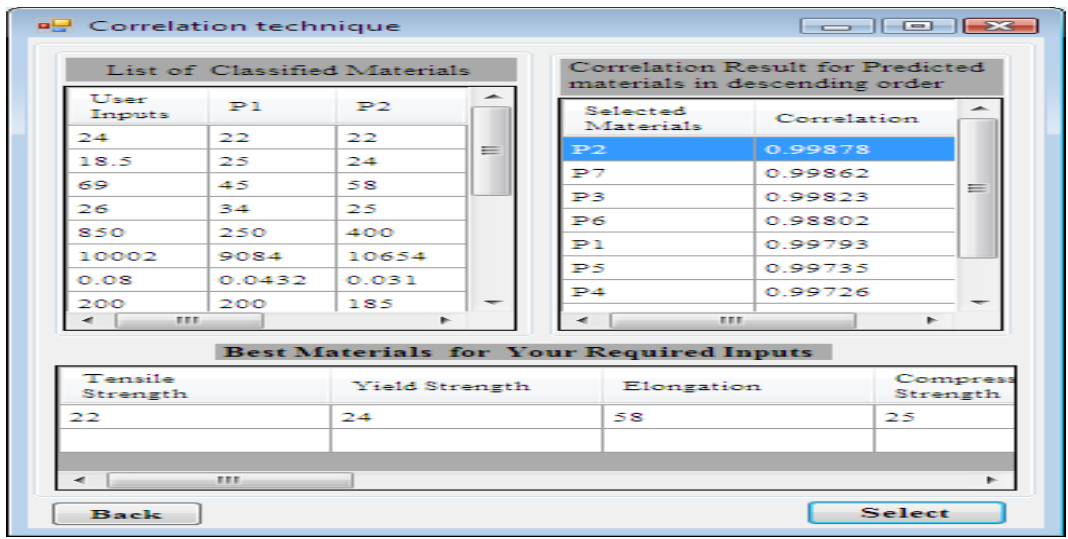

**Fig 4** : Prototype Software module for correlation measure.

The degree of similarity between the input design requirements and the materials data sets in the selected class are measured with numeric attribute values. The materials data sets, whose degree of similarity is very closure to the input design requirements are selected for further materials selection. The degree of similarity threshold range is determined by the correlation range, $0.997 \leq r \leq +1$. The material data sets having the degree of similarity values lies in the range are considered as possible and optimal materials that are closure to the input design requirements. A material data set having the highest degree of similarity is selected as the desired material that satisfies the input design requirements. The possible material data sets discovered from the huge material data sets are listed in the figure 4.

In the figure 5, user input specifications for design requirements with numerical attribute values are shown. These attribute may be computed by CAD systems or directly by a design engineer.

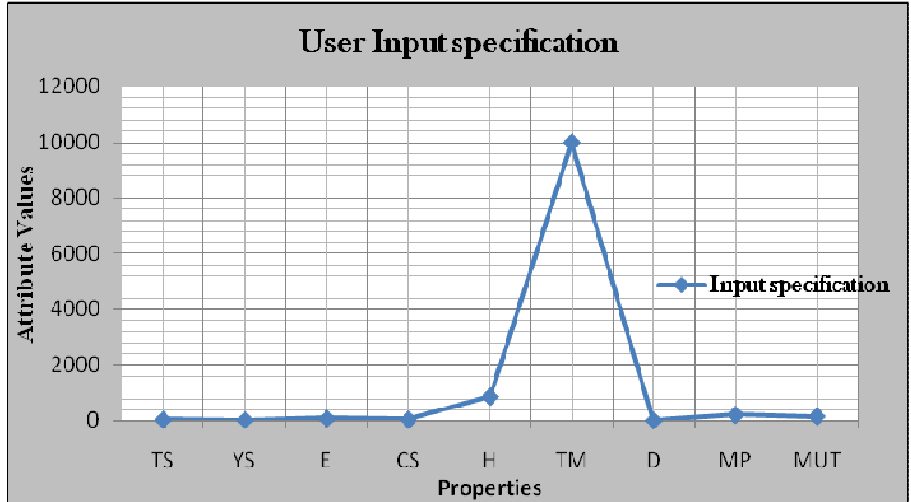

**Fig. 5**: User input specifications

The material data sets selected with correlation measure threshold range  $0.997 \leq r \leq +1$ are shown in figure 6.

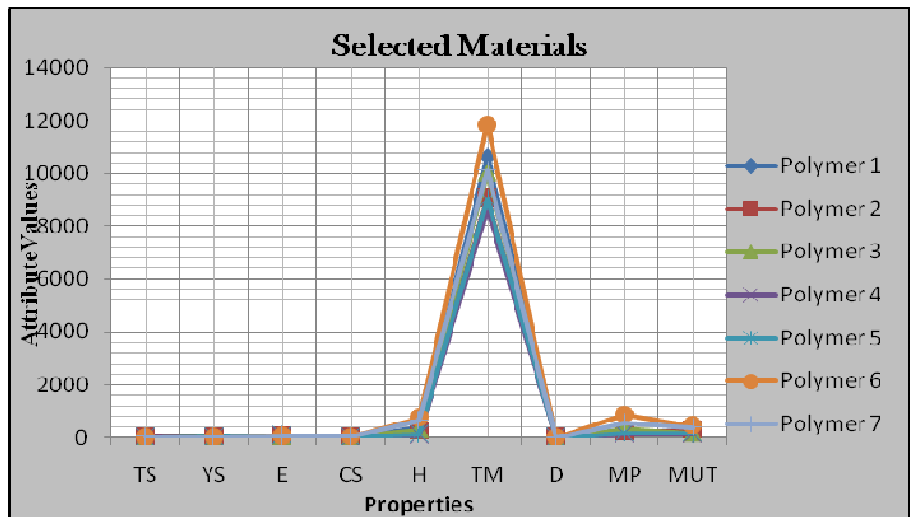

**Fig. 6:** Materials selected from the threshold range  $0.997 \leq r \leq +1$

Selection of an optimal material data from the selected materials is determined by the highest degree of correlation measure. In the figure 7, material say P2 has the highest degree of correlation value compare to other materials ( P1,P3,P4,P5,P6,and P7).Therefore, the material P2 is selected has the optimal material that suits the input design specifications.

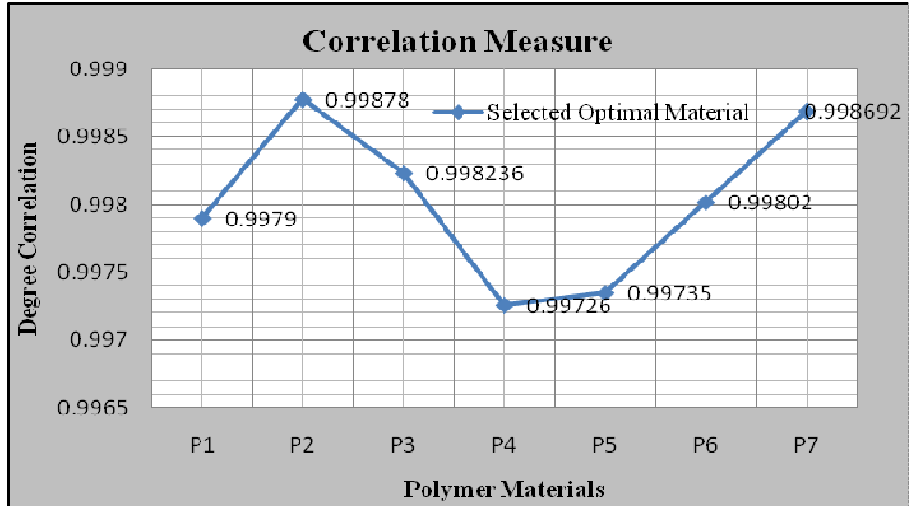
**Fig. 7** : Correlation measures of the selected materials in figure 6.

The user input specifications for design requirements and the selected material numeric attribute values are depicted in the figure 8.

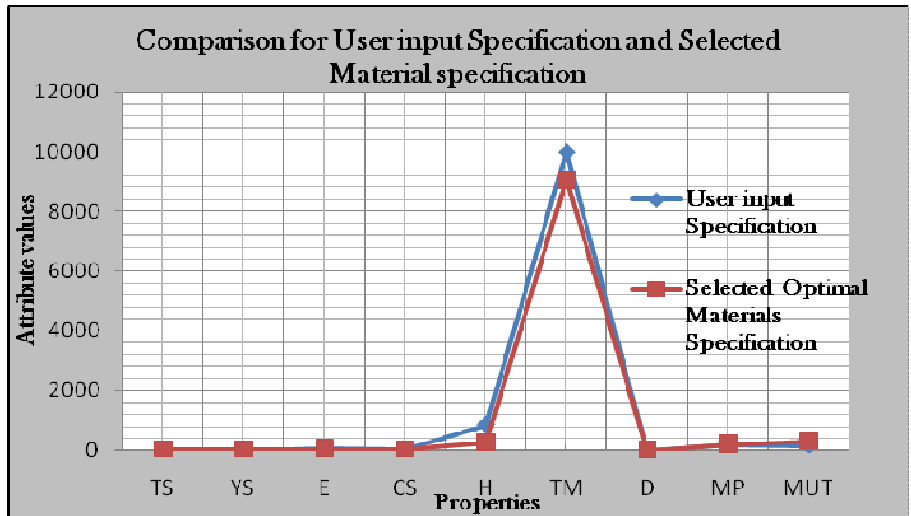
**Fig. 8:** Comparison of user input design requirements and the selected optimal material attribute values.

## 5. CONCLUSIONS AND FUTURE SCOPE

In this paper, an hybrid predictive data mining technique for knowledge discovery from materials database is proposed. Predictive naive Bayesian classifier and Pearson correlation similarity measure are applied to materials informatics for materials classification and selection respectively. The proposed technique for knowledge discovery from materials dataset is successively implemented. The algorithm of the naive Bayesian classifier is applied successively for enabling it to solve classification problems and its outcome is the predicted materials class. With applying Person correlation similarity measure on the outcome of classified material data sets, possible materials are selected. An optimal material that is very closure to the input design requirements is selected by determining the highest degree of similarity measure in the threshold range, $0.997 \leq r \leq +1$, values.

Further, an application of this algorithm can be extended to the classification of engineering materials data sets consisting of both numerical and categorical attribute values using fuzzy system. Performance comparison of this algorithm with other classification algorithms on materials informatics data sets is the future scope of this research.

## ACKNOWLEDGEMENTS

The authors wish to acknowledge the financial support from the University Grant Commission(UGC), INDIA for the Major Research Project **"Scientific Knowledge Discovery Systems (SKDS) For Advanced Engineering Materials Design Applications"** vide reference F.No. 34-99\2008 (SR), 30th December 2008, and also gratefully acknowledge the unanimous reviewers for their kind suggestions and comments for improving this paper.

**Authors**

**Doreswamy** received B.Sc degree in Computer Science and M.Sc Degree in Computer Science from University of Mysore in 1993 and 1995 respectively. After completion of his Post-Graduation Degree, he subsequently joined and served as Lecturer in Computer Science at St.Joseph's College, Bangalore from 1996-1999 and at Yuvaraja's College, a constituent college of University of Mysore from 1999-2002. Then he has elevated to the position Reader in Computer Science at Mangalore University in year 2003. He was the Chairman of the Department of Post-Graduate Studies and Research in Computer Science from 2003-2005 and from 2008-2009 and served at varies capacities in Mangalore University and at present he is the Chairman of Board of Studies in Computer 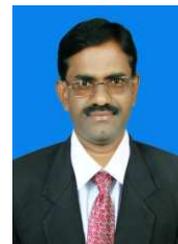 Science of Mangalore University. His areas of research interests include Data Mining and Knowledge Discovery, Artificial Intelligence and Expert Systems, Bioinformatics, Molecular Modeling and Simulation, Computational Intelligence, Nanotechnology, Image Processing and Pattern Recognition. He has been granted a Major Research project entitled "Scientific Knowledge Discovery Systems(SKDS) for advanced Engineering Materials Design Applications" from the funding Agency University Grant Commission, New Delhi, INDIA. He has published about 30 contributed peer reviewed papers at National/International Journals and Conferences. He received SHIKSHA RATTAN PURSKAR for his out standing achievements in the year 2009 and RASTRIYA VIDYA SARAWATHI AWARD for outstanding achievement in chosen field of activity in the year 2010.

**Hemanth K S** received B.Sc degree and MCA degree in the years 2006 and 2009 respectively. Currently working as Project Fellow of UGC Major Research Project and is working towards his Ph.D degree in Computer Science under the guidance of Dr. Doreswamy in the Department of Post-Graduate Studies and Research in Computer Science, Mangalore University.
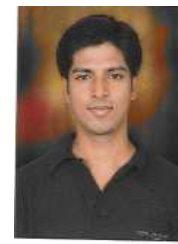